\documentclass[11pt,twoside]{article}

\usepackage{asp2006}
\usepackage{epsf}
\usepackage{lscape}
\newcommand\ho{$h^{-1}$} 

\begin{document}
\title{Megaparsec-Scale Triggers for Star Formation: Clusters
  and Filaments of Galaxies in the Horologium-Reticulum Supercluster} 
\author{Matthew C. Fleenor\altaffilmark{1} \&\ Melanie
  Johnston--Hollitt\altaffilmark{2}}  
\altaffiltext{1}{Physics Group, Roanoke College, 221 College Lane,
  Salem, Virginia}       

\altaffiltext{2}{School of Chemical and Physical Sciences, Victoria University
  of Wellington, PO Box 600, Wellington, New Zealand} 

\begin{abstract} 
Specific indications of star-formation are presented within cluster and
filament galaxies that are members of the Horologium-Reticulum
supercluster (HRS, $z\sim0.06$).  These  
indicators arise from multi-wavelength observations, primarily
emission lines from optical spectroscopy and faint excess from radio
continuum (1.4 GHz) photometry.  HRS galaxies exhibiting current star
formation are consistent with previous studies in that the
star-forming populations organize around megaparsec-scale filament
axes as well as near the cluster core.  Therefore with support from
optical photometry, mechanisms for triggering star formation in these
galaxies are most likely due to merger interactions in lower density
(and lower velocity) environments and possible bursts prior to
stripping within the higher density (and higher velocity) environments.
\end{abstract}

\section{Overview}  
The supercluster environment and its constituent structures have a
role in the processes of galaxy star formation and its associated
triggering mechanisms. 
Mature (low $z$) and dense superclusters ($\delta\sim10$) like the
Horologium-Reticulum supercluster (HRS) of galaxies are the products
of the formation of rich galaxy clusters, and they are also dependent
on the evolution of mature voids ($R_{\rm{void}}\sim10$\ho\ Mpc).  
By using specific examples
from the HRS, the purpose is to present some of the recent results
in galaxy supercluster studies as they relate to star formation in
galaxies. 

The HRS is one of the largest supercluster complexes of galaxies in
the low $z$ universe \citep[$z\sim0.06$,][]{fle05}.  A rich dataset
comprised primarily of optical spectroscopy and
radio continuum photometry over small \citep[e.g., $\leq4$
  deg$^2$,][]{joh08} and wide \citep[$\geq300$ deg$^2$,][]{fle06b}
areas on the sky has provided the context for examining star 
formation on a variety of scales.  In fact, knowledge of the
the supercluster environment via such a comprehensive dataset provides
a key in connecting the specific indications of star formation with
their respective mechanism(s).  The presentation of the results will
follow an order-of-magnitude 
approach in obtaining perspectives of the supercluster environment on
the scales of 100, 10, and 1 megaparsecs.  

\section{``Supercluster--Void Network,'' on the scale of 100 megaparsecs}
It is well-known that superclusters of galaxies are regions
of higher galaxy density and are populated by a larger
percentage of quiescent, early-type galaxies
\citep[e.g.,][Tab.\ 1]{ein07}, which follows from a general
understanding of the morphology-density relation extended out to
galaxy superclusters \citep{got03,gra04}.  Therefore as a whole,
star forming galaxies are seemingly less-likely to reside within
superclusters when compared to field galaxies.  However running
counter to the suppression of star-formation in galaxies due 
to dense environments, some studies find an enhancement of star
formation associated with galaxy filaments that stretch on the
order of 10--50\ho\ Mpc \citep[e.g.,][]{por08,pan08}.  One potential cause for
this enhancement is thought to occur due to an increase in
low-velocity galaxy mergers or recent tidal interaction that occur
more readily in gas-rich, poorer groups. 

The increase in galaxy density within superclusters is not uniform,
where clusters and filaments with higher density are contrasted against
voids of lower density.  Figure 1 shows the HRS survey volume as
presented with the interactive, 3D visualization tool, GyVe
\citep{mil06}.  Specifically, intercluster galaxies 
are shown by the filled gray spheres, galaxy clusters by the filled
cylinders, and mature voids ($R_{\rm{void}}\sim10$\ho\ Mpc) with
numbers.  These mature voids are completely absent of 
the fairly bright galaxies ($b_{\rm{J}}<17.5$) selected for our
survey \citep{fle05}, which was carried out in conjunction with the
six-degree field galaxy survey \citep[6dFGS,][]{jon04}.  Particular to
the case of star formation, filaments of galaxies that stretch on the
order of 10--50\ho\ Mpc are detected on the peripheries of mature
voids \citep[$1<R/R_{\rm{void}}<1.5$,][]{fle06b}.  

\begin{figure}[h]
 \epsscale{0.8}
\plotone{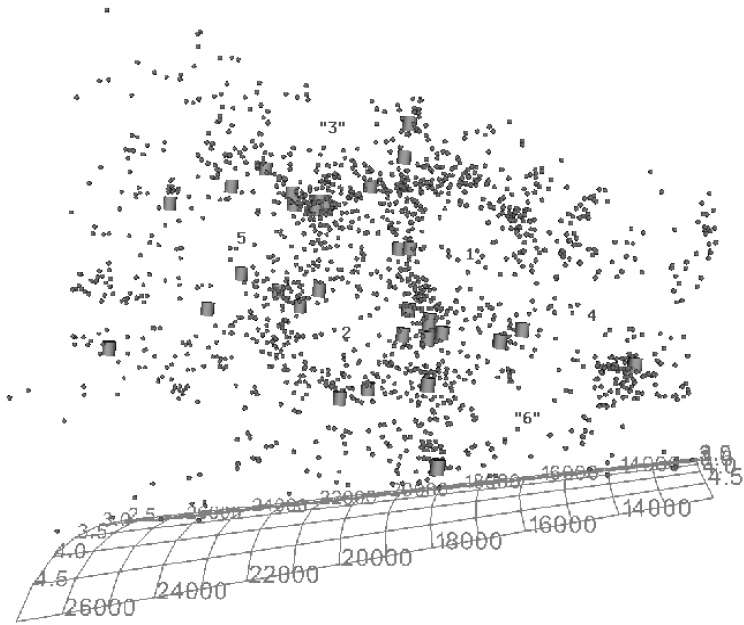}
 \caption{Preferred
  viewing angle snapshot of the 6dF survey sample as taken from
  the GyVe software.  The vantage point is almost
  equivalent to a $\delta (\rm{deg}) - {\it cz} (\rm{km \, s}^{-1})$ 
  plot, where filled spheres are galaxies and filled cylinders are
  galaxy clusters. Numbers show the centers of the mature voids in the
  survey volume.} 
\end{figure}

\section{Filaments of Galaxies, on the scale of 10 megaparsecs}
Guided by the understanding that merger interactions are potentially
heightened in supercluster filaments and the observation that void
peripheries 
contain such filamentary structures in the HRS, galaxies located at
the void periphery are examined for enhanced star formation and 
for understanding the mechanism(s) provoking the enhancement.
Similarly, \citet{cec08} used Sloan and 2dF observations of galaxies 
in `void walls' to show that there are increases in star
formation occurring in bluer galaxies across a wide range in both
luminosity and local density.  Ceccareli et al.\ attribute the
modulation in star formation within these galaxies to the differing
evolutionary history that void galaxies undergo (i.e., less disruptive)
when compared with normal field galaxies.

To quantify the amount of star formation taking place in galaxies
located within $1<R/R_{\rm{void}}<1.5$ (i.e., the void rim), the
H$\alpha$ emission line equivalent widths (EW) were measured from the
6dFGS spectra.  Galaxies are classified as ``active star formers'' when
they display an EW(H$\alpha$) $>10$\AA\ \citep[This value correlates closely
with the 2dFGRS $\eta>0$ that also indicates star
formation,][]{mad02}.  When comparing the ratios of star-forming and 
passive galaxies in void rims to those in Einasto et al., we
find that active star-formation is elevated in bright ($M \leq M^*$)
galaxies by $\sim$30\%\ above normal levels in the supercluster
(210 of 417 total 6dF galaxies).   

To gain a better understanding of the characteristics of the active
star-forming galaxies, Figure 2 shows a representative color-magnitude
relation (CMR) for void rim galaxies in the HRS.  Open circles
represent galaxies in the void rim and filled circles represent
galaxies that have measurements consistent with active star formation.
While the $b_{\rm{J}}$ and $r_{\rm{F}}$ photometric measures are
subject to modest errors ($\pm0.3$mags), most of these star forming
galaxies are above $M_{b_{\rm{J}}}^* \leq -19.3$ \citep[from][in the
  2dFGRS]{nor02}.  Therefore, our results are 
consistent with previous studies in showing a range of luminosities
undergoing active star formation, but they do not seem tightly
correlated with explicitly bluer populations.

\begin{figure}[h]
 \epsscale{0.7}
 \plotone{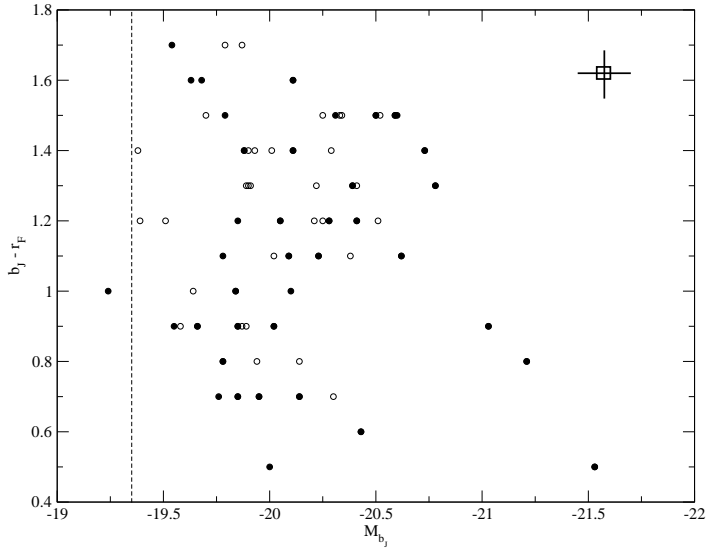}
 \caption{Representative CMR, $b_{\rm{J}} - r_{\rm{F}}$
   vs.\ $M_{b_{\rm{J}}}$, for the 6dF void rim galaxies in voids 1 and
   2 ($1 < R/R_{\rm{void}} < 1.5$).  Open circles represent galaxies
   in the void rim and filled circles represent galaxies that have
   measurements consistent with active star formation.  Quantized
   color effect due to estimated uncertainty in the brightness, where
   error box provides $3\sigma$ errors.}  
\end{figure}

Since the CMR does not show a definite color segregation for the active
star-forming galaxies, photometric images are examined for signature
signs of merging activity (e.g., faint near neighbors, asymmetric or
warped disks, etc.).  Of the active star-forming galaxies in the 6dF void
rims, approximately half ($42\pm5$\%) contain some signature of
interaction.  Figure 3 shows representative examples of 
closely-neighbored, star-forming galaxies, often between brighter
star-forming galaxies and faint bluer galaxies ($M \geq M^*$,
$B-R\sim1.0$) from the total sample of 210 galaxies.  The moderate
density environment of filaments combined with the low velocities
consistent with void evolution seem to suggest that conditions are
favorable for tidal and merging interactions.   

\begin{figure}[h!]
 \epsscale{0.4}
 \plottwo{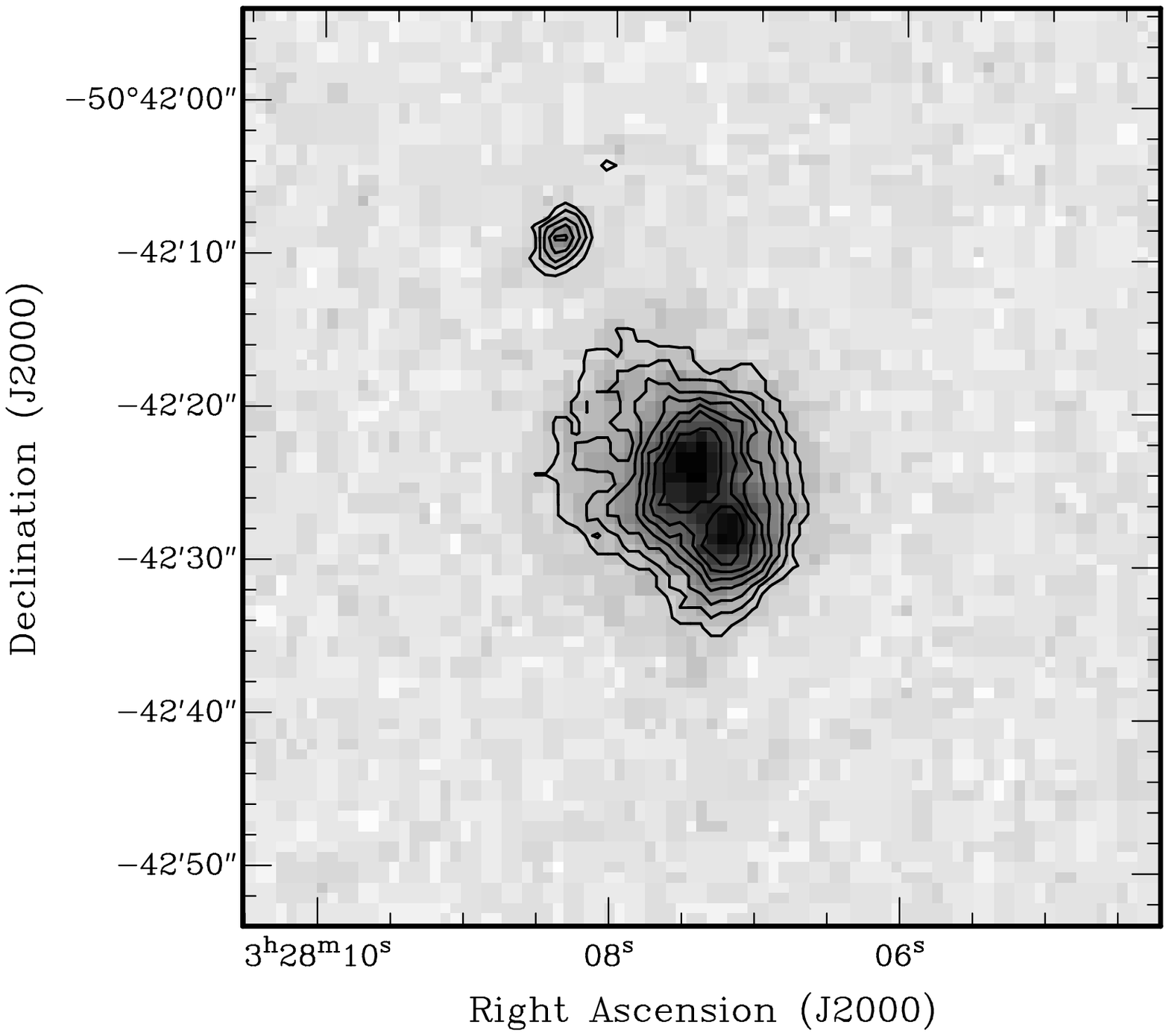}{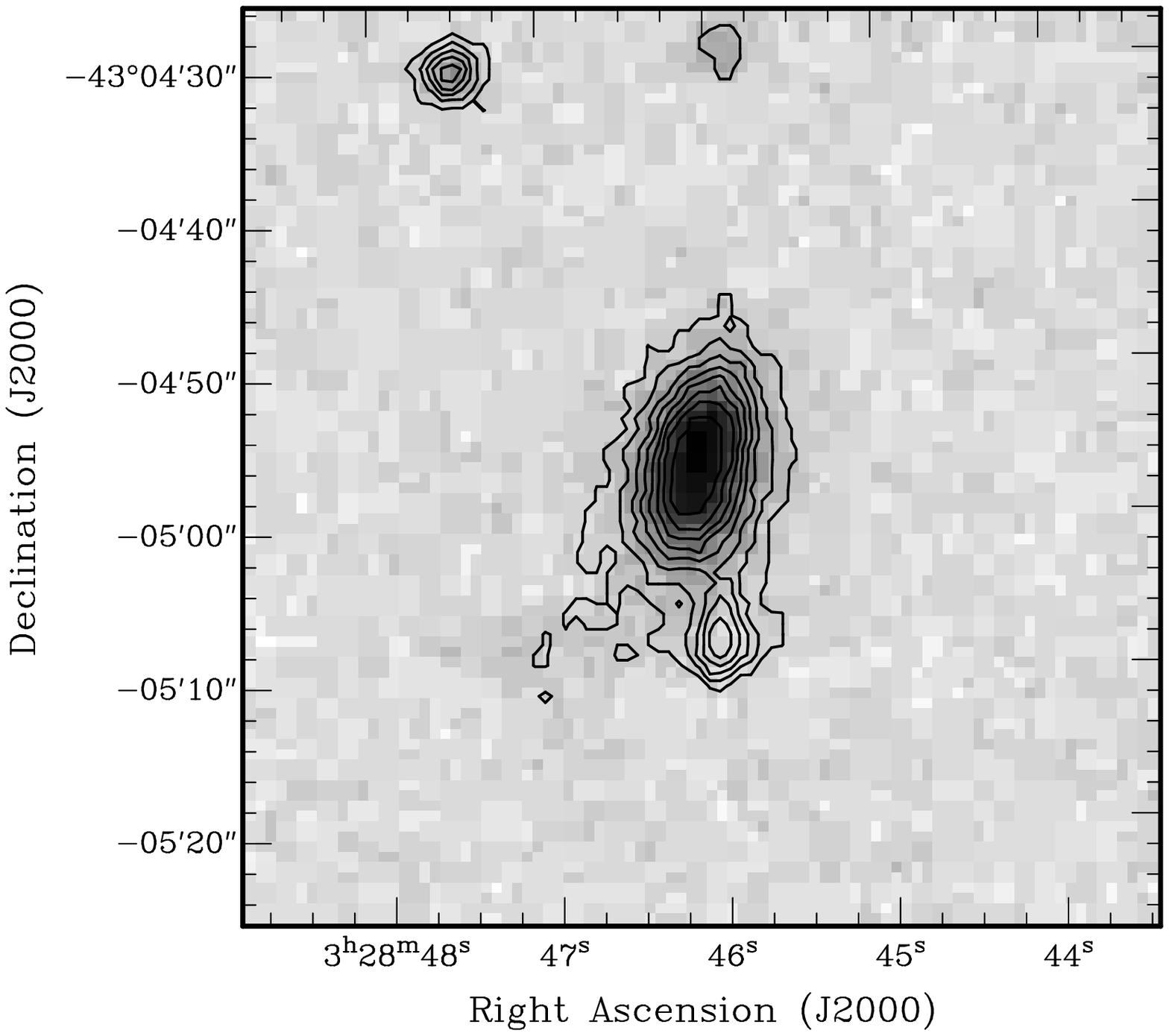}
 \plottwo{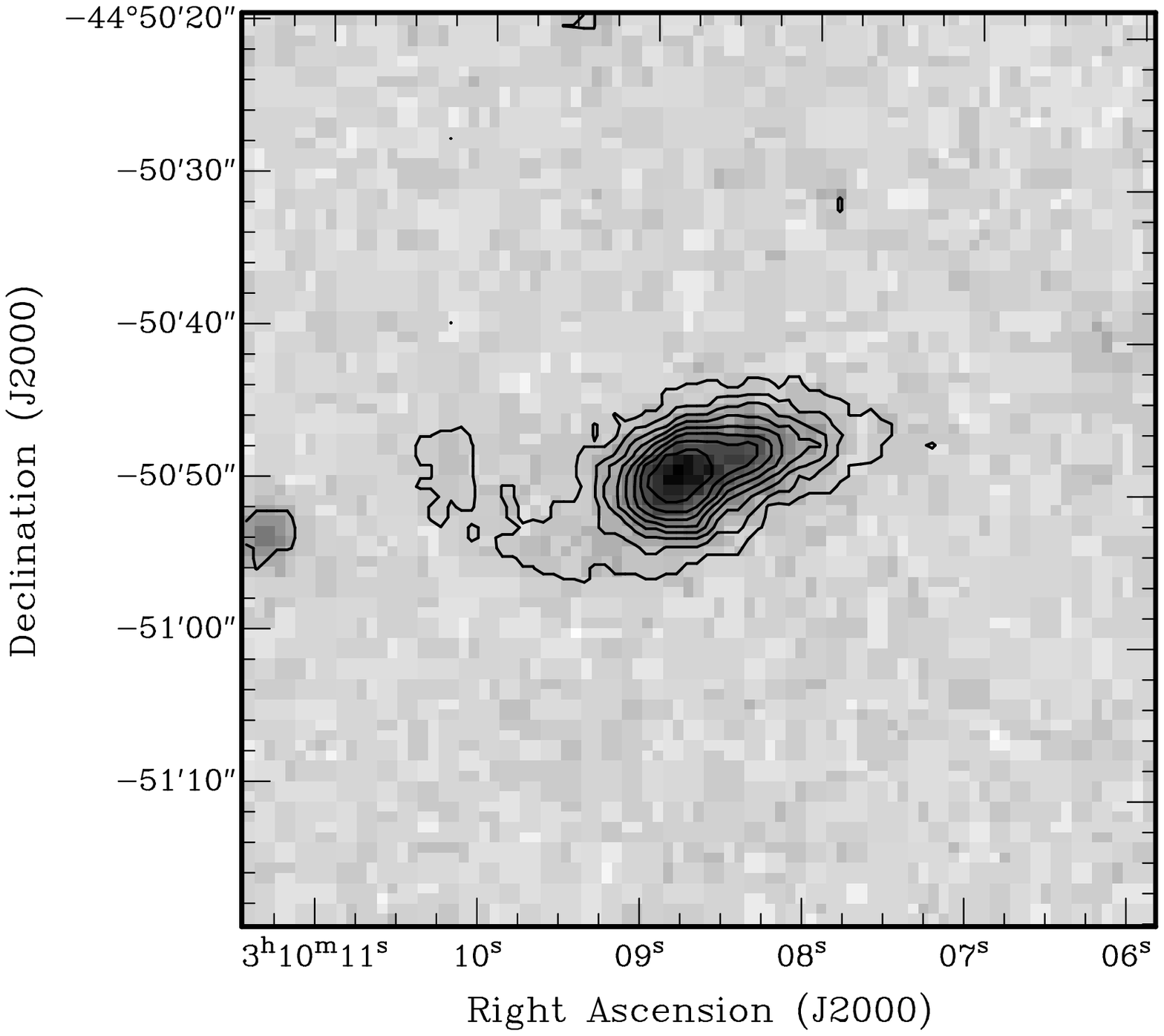}{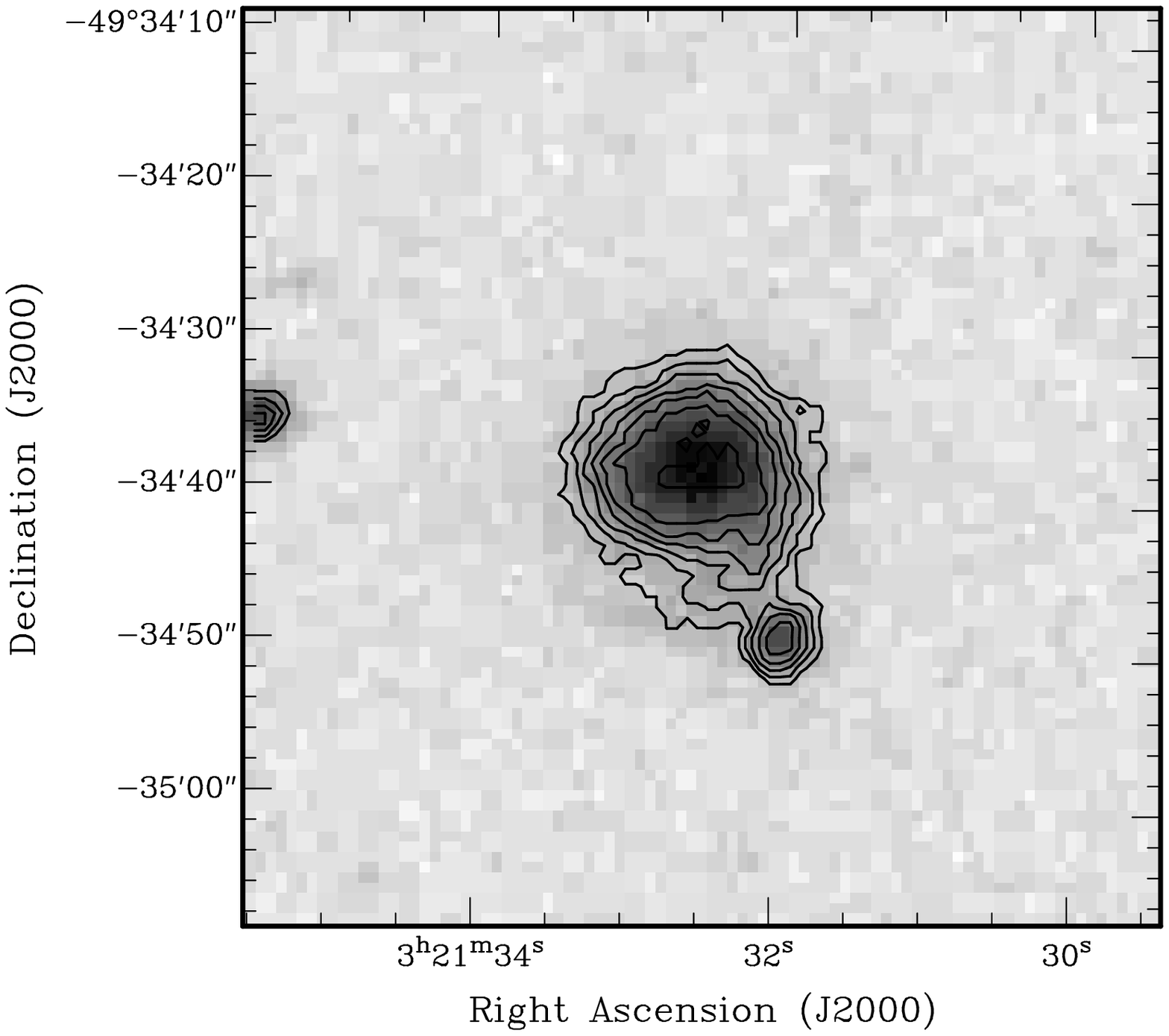}
 \caption{ Representative sample of types of star-forming galaxies
   found in HRS void rims. Images are shown as SuperCOSMOS grayscale
   optical red images ($r_{\rm{F}}$) are overlayed with blue color
   contours ($b_{\rm{J}}$ at 10\%\ increases).  Moving clockwise from
   top left:  (a) definite merger with two distinct nuclei; (b)
   extremely blue galaxy interacting companion (no detection in the
   red image); (c) warped disk indicated by inner and outer contours
   twisted toward possible merging companion; (d) asymmetric disk with
   potential companion.  Postage stamps are $1' \times 1'$, which is
   equivalent to $\sim75$ kpc on a side at HRS distances.} 
\end{figure}

\section{Galaxy Clusters, scales of 1 megaparsec}   
With detailed information of the HRS filament--cluster network
\citep{fle06b}, plus recent results linking star formation to the
infall region around galaxy clusters along filament axes
\citep[e.g.,][]{por08}, the view of the HRS is  
now restricted to the periphery of the rich galaxy cluster, A3158.
Specifically, A3158 is a member of a 10\ho\ Mpc intercluster filament
that forms a large portion of the southern HRS \citep{fle06b}.
Furthermore, the 1.4 GHz luminosity function of A3158 at radio 
wavelengths shows an apparent suppression of AGN-type emission at higher
radio power when compared with representative galaxy clusters,
which is interpreted as a cluster in the later stages of 
merging \citep{ven00,joh08}.  In combination with the suppression of
emission at higher radio power around the cluster, Johnston-Hollitt et al.\
also find a significant excess of low-power radio galaxies at 1.4GHz (i.e., 
$5.4\times10^{21}\leq L_{\rm{1.4GHz}} \leq6.4\times10^{22}$ W
Hz$^{-1}$) seemingly aligned with the A3158 intercluster filament
axis that also contains two, bright cD galaxies \citep[see
  Fig.\ 10,][]{joh08}.  Such values of radio emission are often
associated with star-formation occurring at 
modest rates, SFR$\sim2M_{\odot}$ yr$^{-1}$ \citep{yun01}. 

Having identified a potential population of star-forming galaxies
using statistical measures (i.e., radio source counts and luminosity
function), the characteristics of these galaxies are examined to
better understand the mechanism(s) responsible for the burst of
star-formation.   While it has recently been observed that galaxies
along filament axes undergo a burst of star formation at larger
distances \citep[$r_{\rm{gx}}\sim$2\ho\ Mpc,][]{por08}, many of the
low-power   radio galaxies in A3158 are located at projected distances
of $\leq1.0$\ho\ Mpc (less than the virial radius).  So deep within the
cluster core, it is expected that star-formation mechanisms must
coincide with higher velocities caused by the gravitational potential.
Coupled with the previous observations of a relatively large
axi-symmetric X-ray temperature distribution \citep{oht01}, which
also aligns with the radio galaxy--intercluster filament axis,
stripping-induced star formation within the fainter and bluer
populations of radio galaxies is a plausible mechanism.  Some radio
galaxies located at these distances also exhibit optical emission
($[\rm{OII}]$ and H$\beta$, \citealp{kat98}) also consistent with a
star-forming population \citep[but is not required, see][]{owe05}.
ICM stripping is also observed in detail for similar cluster galaxies
(bluer and fainter) at equivalent distances, where galaxy
transformations are apparently occurring \citep[e.g.,][]{cro05}. 

\section{Conclusions}   
A rich, multi-wavelength dataset reveals the HRS to contain a variety
of filamentary substructures that seem to facilitate star-formation in
member galaxies.  Specifically, statistical measures based on optical (EWs)
and radio (source counts) datasets indicate the presence of enhanced
star-forming populations in both lower density (void periphery) and
higher density (cluster) environments.  In both cases, galaxy
filaments apparently aid the density-specific mechanisms.  Filaments
are effective organizers of material to provide a low-velocity
environment with increased density, where star formation mechanisms
like mergers and tidal interactions can operate effectively (Fig. 3).
In higher density environments with increased infall velocities,
galaxy filaments appear to be conduits for ICM interactions
(like stripping) within fainter and bluer members \citep{joh08}.

\acknowledgements 
MCF thanks J.A.\ Rose and M.E.\ Potts for their collaborative efforts.
MCF also recognizes generous funding from the Roanoke College internal
research grant program.

\end{document}